\documentclass[paper,notoc,nohyper]{JHEP}
\usepackage{epsfig}
\usepackage{amsbsy} 

\def\T{{\cal T}}
\def\C{{\cal C}}
\def\L{{\cal L}}

\def\F{{\cal F}}
\def\G{{\cal G}}

\def\Bfin{{\rm Box^6} (u, t) }
\def\Bubl{{\rm Bub}(s)}

\def\AAA{C_1}
\def\BBB{C_2}
\def\CCC{C_3}

\def\Lx{X}
\def\Ly{Y}
\def\Ls{S}
\def\Poles{{\cal P}oles}
\def\Finite{{\cal F}inite}
\def\Libx{{\rm Li}_2(x)}
\def\Liby{{\rm Li}_2(y)}
\def\Licx{{\rm Li}_3(x)}
\def\Licy{{\rm Li}_3(y)}
\def\Lidx{{\rm Li}_4(x)}
\def\Lidy{{\rm Li}_4(y)}
\def\Lidz{{\rm Li}_4(z)}

\def\ttouu{\frac{t^2}{u^2}}

\def\utoss{\frac{ut}{s^2} }
\def\one{}
\def\tou{\frac{t}{u}}

\def\ttoss{\frac{t^2}{s^2}}

\def\CA{C_A}

\def\NF{N_F}

\def\C{{\cal C}}
\def\D{{\cal D}}
\renewcommand\O[1]{{\cal O}\left(#1\right)}
\def\as{\ensuremath{\alpha_{s}}}
\def\a0{\alpha_0}

\def\Re{\mathop{\rm Re}}

\def\beq{\begin{equation}}
\def\eeq{\end{equation}}

\def\beqn{\begin{eqnarray}}
\def\eeqn{\end{eqnarray}}

\def\lq{\left[}
\def\rq{\right]}

\def\({\left(}
\def\){\right)}

\def\ket#1{|{#1}\rangle}
\def\bra#1{\langle{#1}|}
\def\braket#1#2{\langle #1 |#2 \rangle}
\def\cm{{\cal M}}

\def\MSbar{$\overline{{\rm MS}}$}

\def\bom#1{{\mbox{\boldmath $#1$}}}

\def\fs{\(-\frac{\mu^2}{s}\)^\ep }
\def\ft{\(-\frac{\mu^2}{t}\)^\ep }
\def\fu{\(-\frac{\mu^2}{u}\)^\ep }

\def \ep{\epsilon}

\title{\boldmath Two-loop QCD corrections to gluon-gluon scattering\footnote{
Work supported in part by the UK Particle Physics and Astronomy Research
Council and by the EU Fourth Framework Programme `Training and Mobility of
Researchers', Network `Quantum Chromodynamics and the Deep Structure of
Elementary Particles', contract FMRX-CT98-0194 (DG 12 - MIHT), in part by the
U.S.~Department of Energy under Grant No.~DE-FG02-95ER40896 and in part by the
University of Wisconsin Research Committee with funds granted by the Wisconsin
Alumni Research Foundation. M.E.T. acknowledges financial support from CONACyT
and the CVCP. We thank the British Council and German Academic Exchange Service
for support under ARC project 1050.} } 
\author{ E.~W.~N.~Glover$^a$,
C.~Oleari$^b$ and M.~E.~Tejeda-Yeomans$^a$\\ $^a$Department of Physics, 
University of Durham,  Durham DH1 3LE,  England\\[1mm] $^b$Department of
Physics,  University of Wisconsin, 1150 University Avenue\\ Madison WI 53706, 
U.S.A.\\[1mm] E-mail: \email{E.W.N.Glover@durham.ac.uk},
\email{Oleari@pheno.physics.wisc.edu}, 
\email{M.E.Tejeda-Yeomans@durham.ac.uk}} 
\abstract{ 
We present the $\O{\as^4}$ virtual QCD corrections to gluon-gluon scattering 
due to the interference of tree and two-loop amplitudes.  We work in
conventional dimensional regularisation and give analytic expressions 
renormalised in the \MSbar\ scheme.  The structure of the infrared divergences
agrees with that predicted by Catani while formulae for the finite remainder
are given  in terms of logarithms and  polylogarithms that are real in the
physical region.   These results, together with those previously obtained for
quark-quark and quark-gluon scattering, complete the two-loop matrix elements
needed for the next-to-next-to-leading order contribution to inclusive jet
production at hadron colliders.  
}

\keywords{QCD, Jets, LEP HERA and SLC Physics, NLO and NNLO Computations}
\preprint{{DCTP/01/14}, {IPPP/01/07}, {MADPH-01-1217}, {hep-ph/0102201}}

\begin{document}

\section{Introduction}
\label{sec:intro}

Accurate perturbative calculations beyond leading order in quantum
chromodynamics (QCD) are an  important ingredient in improving our
understanding of jet production  in current and future high energy collider
experiments at the Tevatron and LHC. At present, next-to-leading order
calculations have become standard and  are used to make comparisons with
experimental data.  For example, the next-to-leading order $\O{\as^3}$
predictions for jet production in $p\bar p$ collisions~\cite{EKS,jetrad} based
on the one-loop matrix elements computed by Ellis and Sexton~\cite{ES} have
been successfully compared with a wide variety of experimental observables 
using data from the Tevatron and the CERN S$p\bar{p}$S.   To date these
comparisons have been limited by both experimental and theoretical
uncertainties at the 10\% level. However, improvements in detector technology,
as well as the expected large increases in the luminosity of the colliding
particles, should significantly improve the quality of the experimental data
and 
will require more accurate theoretical calculations either to claim new physics
or to refine our understanding of QCD.

The theoretical prediction may be improved by including the
next-to-next-to-leading order perturbative predictions.  This has the effect of
(a) reducing the renormalisation scale dependence and (b) improving the
matching of the parton level theoretical jet algorithm with the hadron level
experimental jet algorithm, because the jet structure can be modeled by the
presence of a third parton. The improvement in accuracy expected at
next-to-next-to-leading order can be estimated using the renormalisation group
equations together with the existing leading and next-to-leading order
calculations and is at the 1-2\% level for centrally produced jets with a
transverse energy, $E_T$, of around 100~GeV. 

The full next-to-next-to-leading order prediction  requires the knowledge of
the 
two-loop $2 \to 2$ matrix elements as well as the contributions from the
one-loop  $2 \to 3$ and tree-level $2 \to 4$ processes.   At  
large transverse energies, $E_T \gg m_{\rm quark}$,  the quark masses may
be safely neglected and we therefore focus on the scattering of massless
partons. 
Techniques for computing multiparticle tree amplitudes for 
$2\to 4$ processes, and the associated crossed processes,
are well understood.
For example, the helicity amplitudes for the six gluon $gg\to gggg$, four
gluon-two quark  $\bar{q}q\to gggg$, two gluon-four quark
$\bar{q}q\to\bar{q}^\prime q^\prime gg$ and six quark
$\bar{q}q\to\bar{q}^\prime q^\prime\bar{q}^{\prime\prime} q^{\prime\prime}$
have been computed in Refs.~\cite{6g,4g2q,2g4q,6q}. 
Similarly, amplitudes for the one-loop $2 \to 3$ parton sub-processes $gg\to
ggg$, $\bar{q}q\to ggg$, $\bar{q}q\to\bar{q}^\prime q^\prime g$, and
processes related to these by crossing symmetry, are also known and are
available in~\cite{5g,3g2q,1g4q} respectively.

Although the two-loop contribution for gluon-gluon scattering in
$N=4$ supersymmetric models has been known for some time~\cite{bry}, the
evaluation of the two-loop $2\to 2$ contributions for QCD processes has been a 
challenge for the past few years.  This was mainly due to a lack of knowledge
about planar and crossed double box integrals that arise at two-loops.  In the
massless parton limit and in dimensional regularisation, analytic expressions  
for these basic scalar integrals  have now been provided by
Smirnov~\cite{planarA} and Tausk~\cite{nonplanarA}  as series  in
$\ep=(4-D)/2$, where $D$ is the space-time dimension, together with algorithms
for reducing tensor integral to a basis set of known scalar (master)
integrals~\cite{planarB,nonplanarB}.  This makes the calculation of the
two-loop amplitudes for $2 \to 2$ QCD scattering processes possible. 

Following on from the pioneering work of Bern,  Dixon and Ghinculov~\cite{BDG}
who completed the two-loop calculation of physical $2 \to 2$ scattering
amplitudes for the QED  processes $e^+e^- \to \mu^+\mu^-$ and $e^+e^- \to
e^-e^+$, we have studied the $\O{\as^4}$ contributions arising from the
interference of two-loop and tree-level graphs for the QCD processes of
quark-quark~\cite{qqQQ,qqqq,1loopsquare} and quark-gluon~\cite{qqgg}
scattering. In these papers we presented
analytic expressions for the infrared pole structure (that ultimately cancels
against contributions from the $2\to 3$ and $2\to 4$ processes), which agrees
with that anticipated by Catani~\cite{catani},  as well as the
finite remainder. 

To complete the set of two-loop contributions to parton-parton scattering
requires the study of (non-supersymmetric) gluon-gluon scattering. Bern, Dixon and
Kosower~\cite{bdk} were the first to address this process and provided analytic
expressions for the maximal-helicity-violating two-loop amplitude.
Unfortunately, this amplitude vanishes at tree level and does not contribute to
$2 \to 2$ scattering at next-to-next-to-leading order $\O{\as^4}$.  It is
therefore the goal of this paper to provide analytic expressions for the  
$\O{\as^4}$ two-loop corrections to gluon-gluon scattering
\begin{equation} 
g + g \to  g  + g.
\label{eq:gggg}
\end{equation}
As is in Refs.~\cite{qqQQ,qqqq,1loopsquare,qqgg},  we use the \MSbar\
renormalisation  scheme to remove the ultraviolet singularities and
conventional dimensional regularisation, where all external particles are
treated in $D$ dimensions.  We provide expressions for the interference of
tree-level and two-loop graphs. The infrared-pole
structure  agrees with that obtained using Catani's general factorisation
formulae~\cite{catani}.   The finite remainders are the main new results
presented in this paper and we give explicit analytic expressions valid for the
gluon-gluon scattering process in  terms of logarithms and polylogarithms  that
are real in the physical domain.   For simplicity, we decompose our results
according to the powers of the number of colours $N$ and the number of
light-quark flavours $\NF$.

Our paper is organised as follows.  We first establish our notation in
Sec.~\ref{sec:notation}. Analytic expressions for the interference of the
two-loop and tree-level amplitudes are given in Sec.~\ref{sec:two}.   In
Sec.~\ref{subsec:poles} we adopt the notation used in Ref.~\cite{catani}, to
isolate the infrared singularity structure of the two-loop amplitudes in the
\MSbar\ scheme in terms of the one-loop bubble integral in $D=4-2\ep$ and the
one-loop box integral in $D=6-2\ep$.  Analytic formulae connecting these
integrals in the various kinematic regions are given in
Appendix~\ref{app:master_int}. We demonstrate that the anticipated singularity
structure agrees with our explicit calculation. The finite $\O{\ep^0}$
is given in
Sec.~\ref{subsec:finite} in terms of logarithms and polylogarithms that have no
imaginary parts.  Finally we conclude with a brief summary of the results in
Sec.~\ref{sec:conc}.

\section{Notation} 
\label{sec:notation}
For calculational purposes, the process we consider is
\begin{equation}
\label{eq:proc}
g (p_1) + g (p_2)  + g (p_3) + g (p_4) \to 0,
\end{equation}
where the gluons are all incoming with light-like momenta, 
satisfying 
$$
p_1^\mu+p_2^\mu+p_3^\mu+p_4^\mu = 0, \qquad p_i^2=0.
$$
The associated Mandelstam variables are given by
\begin{equation}
s = (p_1+p_2)^2, \qquad t = (p_2+p_3)^2, \qquad u = (p_1+p_3)^2, 
\qquad s+t+u = 0.
\end{equation}
The gluons also carry colour indexes, $a_i$, in the adjoint representation.

We work in conventional dimensional regularisation  treating all external 
quark and gluon states in $D$ dimensions  and renormalise the ultraviolet 
divergences in the \MSbar\ scheme.
The renormalised four point amplitude in the \MSbar\  scheme can be written
\beqn
\ket{\cm}&=& 4\pi \as \Biggl [  \ket{\cm^{(0)}}  
+ \left(\frac{\as}{2\pi}\right)  \ket{\cm^{(1)}} 
+ \left(\frac{\as}{2\pi}\right)^2\,
 \ket{ \cm^{(2)}}  + \O{\as^3} \Biggr ],\nonumber \\
\eeqn
where $\as \equiv \alpha_s(\mu^2)$ is the running coupling at renormalisation 
scale $\mu$
and the $\ket{\cm^{(i)}}$ represents the colour-space vector describing the
renormalised
$i$-loop amplitude. The dependence on both renormalisation scale $\mu$ and
renormalisation scheme is implicit.

We denote the squared amplitude summed over spins and colours by
\beq
\braket{\cm}{\cm} = \sum |{\cal M}({g + g \to  g + g })|^2
= \D(s,t,u).
\eeq
which is symmetric under the exchange of $s$, $t$ and $u$.
The function $\D$ can be expanded perturbatively to yield
\beqn
\D(s,t,u) &=& 16\pi^2\as^2 \left[
 \D^4(s,t,u)+\left(\frac{\as}{2\pi}\right) \D^6(s,t,u)
 +\left(\frac{\as}{2\pi}\right)^2 \D^8(s,t,u) +
\O{\as^{3}}\right],\nonumber \\  
\eeqn
where
\beqn
\D^4(s,t,u) &=& \braket{\cm^{(0)}}{\cm^{(0)}} 
\nonumber\\
&=& 16\,V N^2 (1-\ep)^2\(3-\frac{ut}{s^2}-\frac{us}{t^2}-\frac{st}{u^2}\),\\
\D^6(s,t,u) &=& \left(
\braket{\cm^{(0)}}{\cm^{(1)}}+\braket{\cm^{(1)}}{\cm^{(0)}}\right),\\
\D^8(s,t,u) &=& \left( \braket{\cm^{(1)}}{\cm^{(1)}} +
\braket{\cm^{(0)}}{\cm^{(2)}}+\braket{\cm^{(2)}}{\cm^{(0)}}\right), 
\eeqn
where $N$ is the number of colours and $V=N^2-1$.
Expressions for $\D^6$ are given in Ref.~\cite{ES} using dimensional
regularisation to isolate the infrared and ultraviolet singularities.  

In the following sections, we present expressions for the infrared singular 
and finite two-loop contributions to $\D^8$
\begin{equation}
\D^{8\, (2 \times 0)}(s,t,u) =
\braket{\cm^{(0)}}{\cm^{(2)}}+\braket{\cm^{(2)}}{\cm^{(0)}}.
\end{equation} 
We defer the self-interference of the one-loop amplitudes
\begin{equation}
\label{eq:self_interf}
\D^{8\, (1 \times 1)}(s,t,u) = \braket{\cm^{(1)}}{\cm^{(1)}},
\end{equation} 
to a later paper.

As in Refs.~\cite{qqQQ,qqqq,1loopsquare,qqgg}, we use {\tt QGRAF}~\cite{QGRAF} 
to  produce the
two-loop Feynman diagrams to construct $\ket{\cm^{(2)}}$. 
We then project by 
$\bra{\cm^{(0)}}$ and perform the summation over colours and
spins. It should be noted that when summing over the gluon polarisations, 
we ensure
that the polarisations states are transversal (i.e. physical) by the use of 
an axial
gauge
\begin{equation}
\sum_{{\rm spins}} \ep_{i}^{\mu}\ep_{i}^{\nu *} = 
-  g^{\mu \nu} + \frac{n_{i}^{\mu}p_{i}^{\nu} 
+ n_{i}^{\nu}p_{i}^{\mu}}{n_{i} \cdot p_{i}} 
\end{equation}
where $p_{i}$ is the momentum, $\ep_{i}$ is the polarisation vector and 
$n_{i}$ is an arbitrary light-like 4-vector for gluon $i$. 
For simplicity, we choose 
$n_1^{\mu} = p_2^{\mu}$, $n_2^{\mu} = p_1^{\mu}$,
$n_3^{\mu} = p_4^{\mu}$ and $n_4^{\mu} = p_3^{\mu}$.
Finally, the  trace over the Dirac matrices is carried  out in $D$
dimensions using conventional dimensional regularisation. It is then
straightforward to identify the scalar and tensor integrals present  and
replace them with combinations of the basis set  of master integrals using
the  tensor reduction of two-loop integrals described in~\cite{planarB,nonplanarB,AGO3}, based on integration-by-parts~\cite{IBP} and 
Lorentz invariance~\cite{diffeq} identities.   The final result is  a
combination of master integrals in $D=4-2\epsilon$ for which the 
expansions around $\ep = 0$ are given
in~\cite{planarA,nonplanarA,planarB,nonplanarB,AGO3,AGO2,xtri,bastei3,bastei2}.   

\section{Two-loop contribution}
\label{sec:two}
We further decompose the two-loop contributions as a sum of two terms
\beq
\D^{8 \, (2\times 0)}(s,t,u)
 = \Poles(s,t,u)+\Finite(s,t,u).
\eeq 
$\Poles$ contains infrared singularities that will be  analytically
canceled by those occurring in radiative processes of the
same order (ultraviolet divergences are removed by renormalisation).
$\Finite$ is the remainder which is finite as $\ep \to 0$.

\subsection{Infrared Pole Structure}
\label{subsec:poles}
Following the procedure outlined in Ref.~\cite{catani}, we can write the
infrared pole structure of the two loop contributions renormalised in the 
\MSbar\ scheme in terms of the tree and unrenormalised one-loop amplitudes,
$\ket{\cm^{(0)}}$ and $\ket{\cm^{(1,un)}}$ respectively, as
\begin{eqnarray}
\label{eq:poles}
\Poles = 2 \Re \Biggl[ &&  -\frac{1}{2}\bra{\cm^{(0)}} {\bom I}^{(1)}(\ep){\bom I}^{(1)}(\ep) \ket{\cm^{(0)}}
  -\frac{2\beta_0}{\epsilon}  
\,\bra{\cm^{(0)}} {\bom I}^{(1)}(\ep)  \ket{\cm^{(0)}}
 \nonumber\\
&& 
+\,  \bra{\cm^{(0)}} {\bom I}^{(1)}(\ep)  \ket{\cm^{(1,un)}}
 \nonumber\\
&& 
+
e^{-\ep \gamma } \frac{ \Gamma(1-2\ep)}{\Gamma(1-\ep)} 
\left(\frac{\beta_0}{\epsilon} + K\right)
\bra{\cm^{(0)}} {\bom I}^{(1)}(2\ep) \ket{\cm^{(0)}}\nonumber \\
&&+ \, \bra{\cm^{(0)}}{\bom H}^{(2)}(\ep)\ket{\cm^{(0)}} \Biggr]
\end{eqnarray}
where the Euler constant $\gamma=0.5772\ldots$.
The first coefficient of the QCD beta function, 
$\beta_0$,  for $N_F$ (massless) quark flavours is
\beq
\beta_0 = \frac{11 \CA - 4 T_R \NF}{6},  \quad\quad \CA = N, 
\qquad T_R = \frac{1}{2},
\eeq
and the constant $K$ is
\beq
K = \left( \frac{67}{18} - \frac{\pi^2}{6} \right) \CA - 
\frac{10}{9} T_R \NF.
\eeq
Note that the unrenormalised one-loop amplitude $\ket{\cm^{(1,un)}}$ is what
is obtained by direct Feynman diagram evaluation of the one-loop graphs.

It is convenient to decompose $\ket{\cm^{(0)}}$ and $\ket{\cm^{(1,un)}}$ in
terms 
of $SU(N)$ matrices in the fundamental representation, $T^{a}$, 
so that the tree amplitude may be written as~\cite{coldeca,coldecb,coldecc}
\beq
\label{eq:tree}
\ket{\cm^{(0)}} = \sum_{P(2,3,4)} {\rm Tr}\(T^{a_1}T^{a_2}T^{a_3}T^{a_4}\)
{\cal A}^{{\rm tree}}_4(1,2,3,4),
\eeq
while the one-loop amplitude has the form~\cite{coldec1,coldec2}
\beqn
\label{eq:loop}
\ket{\cm^{(1,un)}} &=& N\sum_{P(2,3,4)} {\rm
Tr}\(T^{a_1}T^{a_2}T^{a_3}T^{a_4}\) 
{\cal A}^{[1]}_{4;1}(1,2,3,4)\nonumber \\
&+&  \sum_{Q(2,3,4)}{\rm Tr}\(T^{a_1}T^{a_2}\){\rm Tr}\(T^{a_3}T^{a_4}\)
{\cal A}^{[1]}_{4;3}(1,2,3,4)\nonumber \\
&+& \NF \sum_{P(2,3,4)} {\rm Tr}\(T^{a_1}T^{a_2}T^{a_3}T^{a_4}\)
{\cal A}^{[1/2]}_{4;1}(1,2,3,4).
\eeqn
In these expressions $\sum_{P(2,3,4)}$ runs over the 6 permutations of
indices of gluons $2$, 3 and 4 while $\sum_{Q(2,3,4)}$ includes the three
choices of pairs of indices, as it is further detailed in Eq.~(\ref{eq:c_i}).
We note that the tree subamplitudes are further related by cyclic and
reflection properties as well as by the dual Ward
identity~\cite{coldecb,Ward} and more general identities~\cite{other}, while
the subleading-colour loop amplitudes ${\cal A}^{[1]}_{4;3}$ are related to
the leading-colour amplitudes ${\cal A}^{[1]}_{4;1}$~\cite{coldec1,coldec2}.  Some of
these relationships are made explicit using an alternative basis in terms of
$SU(N)$ matrices in the adjoint representation~\cite{adjoint}.

To evaluate Eq.~(\ref{eq:poles}) we find it convenient to express 
$\ket{\cm^{(0)}}$ and $\ket{\cm^{(1,un)}}$ as nine-dimensional vectors in colour
space
\beqn
\ket{\cm^{(0)}}&=&\left(
\T_1,  \
\T_2, \
\T_3, \
\T_4,  \
\T_5,  \
\T_6,  \ 
0,\
0,\
0 \right)^T,\\
\ket{\cm^{(1,un)}}&=&\left(
\L_1,  \
\L_2,  \
\L_3,  \
\L_4,  \
\L_5,  \
\L_6,  \ 
\L_7,  \
\L_8,  \
\L_9  \right)^T,
\eeqn
where $()^T$ indicates the transpose vector. Here the $\T_i$ and $\L_i$ are the
components of $\ket{\cm^{(0)}}$ and $\ket{\cm^{(1,un)}}$
in the colour space spanned by the (non-orthogonal) basis 
\begin{eqnarray}
\label{eq:c_i}
\C_1 &=& {\rm Tr}\(T^{a_1}T^{a_2}T^{a_3}T^{a_4}\),\nonumber \\
\C_2 &=& {\rm Tr}\(T^{a_1}T^{a_2}T^{a_4}T^{a_3}\),\nonumber \\
\C_3 &=& {\rm Tr}\(T^{a_1}T^{a_4}T^{a_2}T^{a_3}\),\nonumber \\
\C_4 &=& {\rm Tr}\(T^{a_1}T^{a_3}T^{a_2}T^{a_4}\),\nonumber \\
\C_5 &=& {\rm Tr}\(T^{a_1}T^{a_3}T^{a_4}T^{a_2}\),\nonumber \\
\C_6 &=& {\rm Tr}\(T^{a_1}T^{a_4}T^{a_3}T^{a_2}\),\nonumber \\
\C_7 &=& {\rm Tr}\(T^{a_1}T^{a_2}\){\rm Tr}\(T^{a_3}T^{a_4}\),\nonumber \\
\C_8 &=& {\rm Tr}\(T^{a_1}T^{a_3}\){\rm Tr}\(T^{a_2}T^{a_4}\),\nonumber \\
\C_9 &=& {\rm Tr}\(T^{a_1}T^{a_4}\){\rm Tr}\(T^{a_2}T^{a_3}\).
\end{eqnarray}
The tree and loop amplitudes $\T_i$ and $\L_i$ are directly obtained in terms
of ${\cal A}^{\rm tree}_4$, ${\cal A}^{[1]}_{4;1}$, ${\cal A}^{[1]}_{4;3}$
and ${\cal A}^{[1/2]}_{4;1}$ by reading off from Eqs.~(\ref{eq:tree}) and
(\ref{eq:loop}). As we will see, the amplitudes themselves are not required
since we compute the interference of tree and loop amplitudes directly.

In the same colour basis, the infrared-singularity operator
$\bom{I}^{(1)}(\ep)$ introduced by Catani~\cite{catani} has the form
\begin{eqnarray}
\lefteqn{\bom{I}^{(1)}(\ep) = - \frac{e^{\ep\gamma}}{\Gamma(1-\ep)}
\left(\frac{1}{\ep^2}+\frac{\beta_0}{N\ep}\right)}
\nonumber \\
&\times&\left(
{\small
\begin{array}{ccccccccc}
N({\tt S}+{\tt T}) & 0 & 0 & 0 & 0 & 0 & ({\tt T}-{\tt U}) & 0 & ({\tt S}-{\tt U}) \\
0 & N({\tt S}+{\tt U}) & 0 & 0 & 0 & 0 & ({\tt U}-{\tt T}) & ({\tt S}-{\tt T}) & 0 \\
0 & 0 & N({\tt T}+{\tt U}) & 0 & 0 & 0 & 0 & ({\tt T}-{\tt S}) & ({\tt U}-{\tt S}) \\
0 & 0 & 0 & N({\tt T}+{\tt U}) & 0 & 0 & 0 & ({\tt T}-{\tt S}) & ({\tt U}-{\tt S}) \\
0 & 0 & 0 & 0 & N({\tt S}+{\tt U}) & 0 & ({\tt U}-{\tt T}) & ({\tt S}-{\tt T}) & 0 \\
0 & 0 & 0 & 0 & 0 & N({\tt S}+{\tt T}) & ({\tt T}-{\tt U}) & 0 & ({\tt S}-{\tt U}) \\
({\tt S}-{\tt U}) & ({\tt S}-{\tt T}) & 0 & 0 & ({\tt S}-{\tt T}) & ({\tt S}-{\tt U}) & 2N{\tt S} & 0 & 0 \\
0 & ({\tt U}-{\tt T}) & ({\tt U}-{\tt S}) & ({\tt U}-{\tt S}) & ({\tt U}-{\tt T}) & 0 & 0 & 2N{\tt U} & 0 \\
({\tt T}-{\tt U}) & 0 & ({\tt T}-{\tt S}) & ({\tt T}-{\tt S}) & 0 & ({\tt T}-{\tt U}) & 0 & 0 & 2N{\tt T}
\end{array}
}
\right)\nonumber \\
\end{eqnarray}
where
\beq
{\tt S} = \fs,\qquad\qquad {\tt T} = \ft, \qquad\qquad {\tt U} = \fu.
\eeq
The matrix $\bom{I}^{(1)}(\ep)$ acts directly as a rotation matrix on
$\ket{\cm^{(0)}}$ 
and $\ket{\cm^{(1,un)}}$ in colour space, to give a new colour vector
$\ket{X}$, equal to $\bom{I}^{(1)}(\ep)\ket{\cm^{(0)}}$,
$\bom{I}^{(1)}(\ep)\bom{I}^{(1)}(\ep)\ket{\cm^{(0)}}$ or
$\bom{I}^{(1)}(\ep)\ket{\cm^{(1,un)}}$.

The contraction of the colour vector $\ket{X}$ with the conjugate tree
amplitude obeys the rule
\beq
\braket{\cm^{(0)}}{X} =\sum_{\rm spins} ~\sum_{\rm colours}
 ~\sum_{i,j=1}^9~ \T_i^* \, X_j \, \C_i^*\, \C_j.
\eeq
In evaluating these contractions, we typically encounter $\sum_{\rm colours}
\C_i^*\,\C_j$ which is given by the $ij$ component of the symmetric
matrix ${\cal
C\!C}$
\begin{equation}
{\cal C\!C} = \frac{V}{16N^2}\left(
\begin{array}{ccccccccc}
\AAA & \BBB & \BBB & \BBB & \BBB & \CCC & NV & -N & NV \\
\BBB & \AAA & \BBB & \BBB & \CCC & \BBB & NV & NV & -N \\
\BBB & \BBB & \AAA & \CCC & \BBB & \BBB & -N & NV & NV \\
\BBB & \BBB & \CCC & \AAA & \BBB & \BBB & -N & NV & NV \\
\BBB & \CCC & \BBB & \BBB & \AAA & \BBB & NV & NV & -N \\
\CCC & \BBB & \BBB & \BBB & \BBB & \AAA & NV & -N & NV \\
NV & NV & -N & -N & NV & NV & N^2 V & N^2 & N^2 \\
-N & NV & NV & NV & NV & -N & N^2 & N^2 V & N^2 \\
NV & -N & NV & NV & -N & NV & N^2 & N^2 & N^2 V
\end{array}
\),
\end{equation}
with 
\begin{equation}
\AAA = N^4 - 3N^2 + 3, \qquad \BBB = 3-N^2, \qquad \CCC = 3+N^2.
\end{equation}
Similarly, we find that the interference of the tree-level amplitudes 
$\sum_{\rm spins} \T_i^* \T_j$ is 
given by ${\cal T\!T}_{ij}$, where

\begin{equation}
\label{eq:TT}
{\cal T\!T} = \frac{64(1-\ep)^2(t^2+ut+u^2)^2}{s^2t^2u^2}\ 
{\cal V}^T {\cal V},
\eeq
and the vector ${\cal V}$ is
\beq
{\cal V} = \(u, \ t, \ s,\  s,\  t,\ u ,\ 0,\ 0,\ 0\),
\eeq
while the interference of the tree-level amplitudes with one-loop
amplitudes
$\sum_{\rm spins} \T_i^* \L_j$ is 
given by ${\cal T\!\!L}_{ij}$, where
\begin{equation}
\label{eq:TL}
{\cal T\!\!L} = {\cal V}^T {\cal W},
\eeq
and the vector ${\cal W}$ is
\beq
{\cal W}= \Big(
\F(s,t),\  \F(s,u) ,\ \F(u,t) ,\ \F(u,t) ,\ \F(s,u) ,\ \F(s,t),\ 
\G, \ \G,\ \G \Big).
\eeq
Here the function $\F(s,t)$ is symmetric under the exchange of $s$ and $t$,
while $\G$ is symmetric under the exchange of any two Mandelstam invariants,
so that
\begin{eqnarray}
\F(s,t) &=& f_1(s,t,u)+f_1(t,s,u),\\
\G &=& f_2(s,t,u)+f_2(s,u,t)+f_2(t,s,u)+f_2(t,u,s)+f_2(u,s,t)+f_2(u,t,s).
\nonumber \\
\end{eqnarray}
Here $f_1$ and $f_2$ are given in terms of the one-loop box integral in $D=6-2\ep$ dimensions and the
one-loop bubble graph in $D=4-2\ep$,
\begin{eqnarray}
f_1(s,t,u) &=& \frac{16N(1-2\ep)}{s^2t^2}\left[2(1-\ep)^2\(s^4+s^3t+st^3+t^4\) + 3(1-5\ep)s^2t^2\right]
{\rm Box}^6(s,t)\nonumber \\
&+& \frac{8\NF(1-2\ep)}{st}\left[(1-\ep)^2 \(s^2+t^2\)+\ep(1+3\ep)st\right] {\rm Box}^6(s,t) \nonumber
\\
&-& \frac{16N(1-\ep)}{s^2t^2u\ep(3-2\ep)}
\left[\(12-22\ep+12\ep^2+2\ep^3\)s^4+\(24-58\ep+50\ep^2-6\ep^3-2\ep^4\)s^3t\right. \nonumber \\
&& \qquad \left.
+\(36-99\ep+93\ep^2-24\ep^3-2\ep^4\)s^2t^2+(1-\ep)\(24-50\ep+23\ep^2\)st^3\right.\nonumber \\
&& \qquad \left.
+4(1-\ep)(1-2\ep)(3-2\ep)t^4\right]
{\rm Bub}(t)\nonumber \\
&+& \frac{16\NF}{st^2u(3-2\ep)}\left[\(4-12\ep+16\ep^2-4\ep^3\)s^3+\(3-10\ep+23\ep^2-8\ep^3\) s^2t 
\right. \nonumber \\
&&\qquad \left. 
+
\(6-15\ep+21\ep^2-8\ep^3\) st^2 + (1-\ep)\(5-6\ep+2\ep^2\)t^3 \right] {\rm
Bub}(t),\\
f_2(s,t,u) &=& \frac{32(1-2\ep)}{u^2}\left[-4(1-\ep)^2st+3(1-5\ep)u^2\right]
{\rm Box}^6(u,t)\nonumber \\
&& + \frac{32(1-\ep)}{\ep su^2}
\left[ 4(1-2\ep)(1-\ep)t^2+(8-17\ep)(1-\ep)ut\right.\nonumber \\
&& \qquad \qquad
\left.+\(6-20\ep+15\ep^2+\ep^3\)u^2
\right] 
{\rm Bub}(s). 
\end{eqnarray}
Series expansions around $\ep = 0$ for the one-loop integrals are given in 
Appendix~\ref{app:master_int}.  

Finally, the last term of Eq.~(\ref{eq:poles}) that involves 
${\bom H}^{(2)}(\ep)$ 
produces only a single pole in $\ep$ and is given by 
\beq
\label{eq:htwo}
\bra{\cm^{(0)}}{\bom H}^{(2)}(\ep)\ket{\cm^{(0)}} 
=\frac{e^{\ep \gamma}}{4\,\ep\,\Gamma(1-\ep)} H^{(2)} 
\braket{\cm^{(0)}}{\cm^{(0)}} \nonumber \\  
\eeq
where the constant $H^{(2)}$ is
\beq
\label{eq:Htwo}
H^{(2)} =  
\left (2\zeta_3+{\frac {5}{3}}+ {\frac {11}{36}}\,\pi^2
\right )N^2
+{\frac {20}{27}}\,\NF^2
+\left (-{\frac {{\pi }^{2}}{18}}-{
\frac {89}{27}}\right ) N \NF -\frac{\NF}{N}, 
\eeq
and $\zeta_n$ is the Riemann Zeta function with $\zeta_2 = \pi^2/6$ and
$\zeta_3 = 1.202056\ldots$  We note that $H^{(2)}$ is renormalisation-scheme
dependent and Eq.~(\ref{eq:Htwo}) is valid in the \MSbar\ scheme.  We also
note that Eq.~(\ref{eq:Htwo}) differs from the corresponding expressions
found in the singularity structure of two-loop 
quark-quark and quark-gluon scattering.  
This is due to double emissions from the gluons.   In fact, we note that $H^{(2)}$ for quark-gluon scattering is the
average of the $H^{(2)}$ for gluon-gluon scattering and quark-quark scattering,
as may be expected by counting the number of different types of radiating
partons.

It can be easily noted that the leading infrared singularity in 
Eq.~(\ref{eq:poles}) is $\O{1/\ep^4}$.  It is a very stringent check on the
reliability of our calculation that  the pole structure obtained by computing
the Feynman diagrams directly and introducing series expansions in $\epsilon$
for the scalar master integrals  agrees with Eq.~(\ref{eq:poles}) through to
$\O{1/\ep}$.   We therefore construct the finite remainder by subtracting
Eq.~(\ref{eq:poles}) from the full result.

\subsection{Finite contributions}
\label{subsec:finite}

The finite two-loop contribution to $\D^8(s,t,u)$ is defined as 
\beq
\Finite(s,t,u) = \D^{8\, (2 \times 0)}(s,t,u) - \Poles(s,t,u),
\eeq
where we subtract the series expansions of both $\D^{8\, (2 \times 0)}(s,t,u)$ 
and $\Poles(s,t,u)$ and set $\ep \to 0$.
As usual, the polylogarithms ${\rm Li}_n(w)$ are defined by
\begin{eqnarray}
 {\rm Li}_n(w) &=& \int_0^w \frac{dt}{t} {\rm Li}_{n-1}(t) \qquad {\rm ~for~}
 n=2,3,4\nonumber \\
 {\rm Li}_2(w) &=& -\int_0^w \frac{dt}{t} \log(1-t).
\label{eq:lidef}
\end{eqnarray} 
Using the standard polylogarithm identities~\cite{kolbig},
we retain the polylogarithms with arguments $x$, $1-x$ and
$(x-1)/x$, where
\begin{equation}
\label{eq:xydef}
x = -\frac{t}{s}, \qquad y = -\frac{u}{s} = 1-x, \qquad z=-\frac{u}{t} = \frac{x-1}{x}.
\end{equation}
For convenience, we also introduce the following logarithms
\begin{equation}
\label{eq:xydef1}
\Lx = \log\left(\frac{-t}{s}\right),
\qquad \Ly = \log\left(\frac{-u}{s}\right),
\qquad \Ls = \log\left(\frac{s}{\mu^2}\right),
\end{equation}
where $\mu$ is the renormalisation scale.

We choose to present our results by grouping terms 
according to the
power of the number of colours $N$ and the number 
of light quarks $\NF$, so that 
\begin{equation}
\label{eq:zi}
Finite(s,t,u) =  V 
\Bigg(
N^4 A + N^2 B + N^3\NF C  + N\NF D  
+ N^2 \NF^2 E  + \NF^2 F
\Bigg),
\end{equation}
where
\begin{eqnarray}
\label{eq:finiteA}
{A}&=&{\Biggl \{ }\Biggl ({48}\,{\Lidx}-{48}\,{\Lidy}-{128}\,{\Lidz}+{40}\,{\Licx}\,{\Lx}-{64}\,{\Licx}\,{\Ly}-{98\over 3}\,{\Licx}\nonumber \\ &&
+{64}\,{\Licy}\,{\Lx}-{40}\,{\Licy}\,{\Ly}+{18}\,{\Licy}+{98\over 3}\,{\Libx}\,{\Lx}-{16\over 3}\,{\Libx}\,{\pi^2}-{18}\,{\Liby}\,{\Ly}\nonumber \\ &&
-{37\over 6}\,{\Lx^4}+{28}\,{\Lx^3}\,{\Ly}-{23\over 3}\,{\Lx^3}-{16}\,{\Lx^2}\,{\Ly^2}+{49\over 3}\,{\Lx^2}\,{\Ly}-{35\over 3}\,{\Lx^2}\,{\pi^2}-{38\over 3}\,{\Lx^2}-{22\over 3}\,{\Ls}\,{\Lx^2}\nonumber \\ &&
-{20\over 3}\,{\Lx}\,{\Ly^3}-{9}\,{\Lx}\,{\Ly^2}+{8}\,{\Lx}\,{\Ly}\,{\pi^2}+{10}\,{\Lx}\,{\Ly}-{31\over 12}\,{\Lx}\,{\pi^2}-{22}\,{\zeta_3}\,{\Lx}+{22\over 3}\,{\Ls}\,{\Lx}+{37\over 27}\,{\Lx}\nonumber \\ &&
+{11\over 6}\,{\Ly^4}-{41\over 9}\,{\Ly^3}-{11\over 3}\,{\Ly^2}\,{\pi^2}-{22\over 3}\,{\Ls}\,{\Ly^2}+{266\over 9}\,{\Ly^2}-{35\over 12}\,{\Ly}\,{\pi^2}+{418\over 9}\,{\Ls}\,{\Ly}+{257\over 9}\,{\Ly}\nonumber \\ &&
+{18}\,{\zeta_3}\,{\Ly}-{31\over 30}\,{\pi^4}-{11\over 9}\,{\Ls}\,{\pi^2}+{31\over 9}\,{\pi^2}+{242\over 9}\,{\Ls^2}+{418\over 9}\,{\zeta_3}+{2156\over 27}\,{\Ls}\nonumber \\ &&
-{11093\over 81}-{8}\,{\Ls}\,{\zeta_3}\Biggr ) {}\,{\ttoss}\nonumber \\ &&
+{}\Biggl ({}-{256}\,{\Lidx}-{96}\,{\Lidy}+{96}\,{\Lidz}+{80}\,{\Licx}\,{\Lx}+{48}\,{\Licx}\,{\Ly}-{64\over 3}\,{\Licx}\nonumber \\ &&
-{48}\,{\Licy}\,{\Lx}+{96}\,{\Licy}\,{\Ly}-{304\over 3}\,{\Licy}+{64\over 3}\,{\Libx}\,{\Lx}-{32\over 3}\,{\Libx}\,{\pi^2}+{304\over 3}\,{\Liby}\,{\Ly}\nonumber \\ &&
+{26\over 3}\,{\Lx^4}-{64\over 3}\,{\Lx^3}\,{\Ly}-{64\over 3}\,{\Lx^3}+{20}\,{\Lx^2}\,{\Ly^2}+{136\over 3}\,{\Lx^2}\,{\Ly}+{24}\,{\Lx^2}\,{\pi^2}+{76}\,{\Lx^2}-{88\over 3}\,{\Ls}\,{\Lx^2}\nonumber \\ &&
+{8\over 3}\,{\Lx}\,{\Ly^3}+{104\over 3}\,{\Lx}\,{\Ly^2}-{16\over 3}\,{\Lx}\,{\Ly}\,{\pi^2}+{176\over 3}\,{\Ls}\,{\Lx}\,{\Ly}-{136\over 3}\,{\Lx}\,{\Ly}-{50\over 3}\,{\Lx}\,{\pi^2}-{48}\,{\zeta_3}\,{\Lx}\nonumber \\ &&
+{2350\over 27}\,{\Lx}+{440\over 3}\,{\Ls}\,{\Lx}+{4}\,{\Ly^4}-{176\over 9}\,{\Ly^3}+{4\over 3}\,{\Ly^2}\,{\pi^2}-{176\over 3}\,{\Ls}\,{\Ly^2}-{494\over 9}\,{\Ly}\,{\pi^2}+{5392\over 27}\,{\Ly}\nonumber \\ &&
-{64}\,{\zeta_3}\,{\Ly}+{496\over 45}\,{\pi^4}-{308\over 9}\,{\Ls}\,{\pi^2}+{200\over 9}\,{\pi^2}+{968\over 9}\,{\Ls^2}+{8624\over 27}\,{\Ls}-{44372\over 81}\nonumber \\ &&
+{1864\over 9}\,{\zeta_3}-{32}\,{\Ls}\,{\zeta_3}\Biggr ) {}\,{\tou}\nonumber \\ &&
+{}\Biggl ({88\over 3}\,{\Licx}-{88\over 3}\,{\Libx}\,{\Lx}+{2}\,{\Lx^4}-{8}\,{\Lx^3}\,{\Ly}-{220\over 9}\,{\Lx^3}+{12}\,{\Lx^2}\,{\Ly^2}+{88\over 3}\,{\Lx^2}\,{\Ly}+{8\over 3}\,{\Lx^2}\,{\pi^2}\nonumber \\ &&
-{88\over 3}\,{\Ls}\,{\Lx^2}+{304\over 9}\,{\Lx^2}-{8}\,{\Lx}\,{\Ly^3}-{16\over 3}\,{\Lx}\,{\Ly}\,{\pi^2}+{176\over 3}\,{\Ls}\,{\Lx}\,{\Ly}-{77\over 3}\,{\Lx}\,{\pi^2}+{1616\over 27}\,{\Lx}\nonumber \\ &&
+{968\over 9}\,{\Ls}\,{\Lx}-{8}\,{\zeta_3}\,{\Lx}+{4}\,{\Ly^4}-{176\over 9}\,{\Ly^3}-{20\over 3}\,{\Ly^2}\,{\pi^2}-{176\over 3}\,{\Ls}\,{\Ly^2}-{638\over 9}\,{\Ly}\,{\pi^2}-{16}\,{\zeta_3}\,{\Ly}\nonumber \\ &&
+{5392\over 27}\,{\Ly}-{4\over 15}\,{\pi^4}-{308\over 9}\,{\Ls}\,{\pi^2}-{20}\,{\pi^2}-{32}\,{\Ls}\,{\zeta_3}+{1408\over 9}\,{\zeta_3}+{968\over 9}\,{\Ls^2}-{44372\over 81}\nonumber \\ &&
+{8624\over 27}\,{\Ls}\Biggr ) {}\,{\ttouu}\nonumber \\ &&
+\,{}\Biggl ({44\over 3}\,{\Licx}-{44\over 3}\,{\Libx}\,{\Lx}-{\Lx^4}+{110\over 9}\,{\Lx^3}-{22\over 3}\,{\Lx^2}\,{\Ly}+{14\over 3}\,{\Lx^2}\,{\pi^2}+{44\over 3}\,{\Ls}\,{\Lx^2}\nonumber \\ & &
-{152\over 9}\,{\Lx^2}-{10}\,{\Lx}\,{\Ly}+{11\over 2}\,{\Lx}\,{\pi^2}+{4}\,{\zeta_3}\,{\Lx}-{484\over 9}\,{\Ls}\,{\Lx}-{808\over 27}\,{\Lx}+{7\over 30}\,{\pi^4}-{31\over 9}\,{\pi^2}\nonumber \\ & &
+{11\over 9}\,{\Ls}\,{\pi^2}-{418\over 9}\,{\zeta_3}-{242\over 9}\,{\Ls^2}-{2156\over 27}\,{\Ls}+{8}\,{\Ls}\,{\zeta_3}+{11093\over 81}\Biggr ) {\utoss}\nonumber \\ & & {}
+{}\Biggl ({}-{176}\,{\Lidx}+{88}\,{\Licx}\,{\Lx}-{168}\,{\Licx}\,{\Ly}-{206\over 3}\,{\Licx}+{206\over 3}\,{\Libx}\,{\Lx}\nonumber \\ & &
+{65\over 6}\,{\Lx^4}-{40\over 3}\,{\Lx^3}\,{\Ly}-{295\over 9}\,{\Lx^3}-{15}\,{\Lx^2}\,{\Ly^2}+{115\over 3}\,{\Lx^2}\,{\Ly}+{29\over 3}\,{\Lx^2}\,{\pi^2}-{670\over 9}\,{\Lx^2}\nonumber \\ & &
-{242\over 3}\,{\Ls}\,{\Lx^2}+{64\over 3}\,{\Lx}\,{\Ly}\,{\pi^2}+{209\over 3}\,{\Lx}\,{\Ly}+{44}\,{\Ls}\,{\Lx}\,{\Ly}-{1811\over 36}\,{\Lx}\,{\pi^2}+{8983\over 27}\,{\Lx}\nonumber \\ & &
+{1870\over 9}\,{\Ls}\,{\Lx}-{18}\,{\zeta_3}\,{\Lx}+{31\over 20}\,{\pi^4}-{361\over 18}\,{\pi^2}-{517\over 18}\,{\Ls}\,{\pi^2}+{1331\over 9}\,{\Ls^2}+{12452\over 27}\,{\Ls}\nonumber \\ & &
+{1543\over 9}\,{\zeta_3}-{129475\over 162}-{44}\,{\Ls}\,{\zeta_3}\Biggr ) {}\,{\one}\Biggr \} + \Biggl \{ u \leftrightarrow t \Biggr \},
\end{eqnarray}
\begin{eqnarray}
\label{eq:finiteB}
{B }&=&{\Biggl \{ }\Biggl ({}-{288}\,{\Lidx}+{480}\,{\Lidy}-{288}\,{\Lidz}+{240}\,{\Licx}\,{\Lx}-{144}\,{\Licx}\,{\Ly}\nonumber \\ &&
+{224}\,{\Licx}+{144}\,{\Licy}\,{\Lx}-{432}\,{\Licy}\,{\Ly}-{224}\,{\Licy}+{48}\,{\Libx}\,{\Lx^2}\nonumber \\ &&
-{224}\,{\Libx}\,{\Lx}-{176}\,{\Libx}\,{\pi^2}+{48}\,{\Liby}\,{\Ly^2}+{224}\,{\Liby}\,{\Ly}-{16}\,{\Lx^4}+{112}\,{\Lx^3}\,{\Ly}\nonumber \\ &&
-{556\over 3}\,{\Lx^3}-{48}\,{\Lx^2}\,{\Ly^2}+{180}\,{\Lx^2}\,{\Ly}-{40}\,{\Lx^2}\,{\pi^2}+{220}\,{\Lx^2}-{32}\,{\Lx}\,{\Ly^3}-{92}\,{\Lx}\,{\Ly^2}\nonumber \\ &&
-{16}\,{\Lx}\,{\Ly}\,{\pi^2}-{376\over 3}\,{\Lx}\,{\Ly}-{16}\,{\Lx}\,{\pi^2}-{80}\,{\Lx}+{96}\,{\zeta_3}\,{\Lx}+{8}\,{\Ly^4}+{292\over 3}\,{\Ly^3}-{32}\,{\Ly^2}\,{\pi^2}\nonumber \\ &&
-{284\over 3}\,{\Ly^2}+{16}\,{\Ly}\,{\pi^2}+{80}\,{\Ly}-{96}\,{\zeta_3}\,{\Ly}+{38\over 5}\,{\pi^4}-{18}\,{\pi^2}\Biggr ) {}\,{\ttoss}\nonumber \\ &&
+{}\Biggl ({}-{576}\,{\Lidx}+{384}\,{\Lidy}-{1152}\,{\Lidz}+{1056}\,{\Licx}\,{\Lx}-{768}\,{\Licx}\,{\Ly}\nonumber \\ &&
+{448}\,{\Licx}+{768}\,{\Licy}\,{\Lx}-{768}\,{\Licy}\,{\Ly}+{896}\,{\Licy}-{192}\,{\Libx}\,{\Lx^2}\nonumber \\ &&
-{448}\,{\Libx}\,{\Lx}-{544}\,{\Libx}\,{\pi^2}-{384}\,{\Liby}\,{\Lx}\,{\Ly}-{896}\,{\Liby}\,{\Ly}-{28}\,{\Lx^4}+{144}\,{\Lx^3}\,{\Ly}\nonumber \\ &&
+{320\over 3}\,{\Lx^3}-{336}\,{\Lx^2}\,{\Ly^2}-{224}\,{\Lx^2}\,{\Ly}-{40}\,{\Lx^2}\,{\pi^2}-{64}\,{\Lx^2}-{32}\,{\Lx}\,{\Ly^3}+{128}\,{\Lx}\,{\Ly^2}\nonumber \\ &&
-{64}\,{\Lx}\,{\Ly}\,{\pi^2}+{1888\over 3}\,{\Lx}\,{\Ly}-{288}\,{\Lx}\,{\pi^2}+{160}\,{\Lx}-{1248}\,{\zeta_3}\,{\Lx}-{240}\,{\Ly^2}\,{\pi^2}-{928}\,{\Ly}\,{\pi^2}\nonumber \\ &&
+{768}\,{\zeta_3}\,{\Ly}+{1216\over 15}\,{\pi^4}-{1912\over 3}\,{\pi^2}-{448}\,{\zeta_3}\Biggr )  {}\,{\tou}\nonumber \\ &&
+{}\Biggl ({}-{384}\,{\Lidy}-{384}\,{\Lidz}+{384}\,{\Licx}\,{\Lx}-{384}\,{\Licx}\,{\Ly}+{384}\,{\Licy}\,{\Lx}\nonumber \\ &&
-{192}\,{\Libx}\,{\Lx^2}-{192}\,{\Libx}\,{\pi^2}-{384}\,{\Liby}\,{\Lx}\,{\Ly}-{8}\,{\Lx^4}-{32}\,{\Lx^3}\,{\Ly}-{176}\,{\Lx^3}\nonumber \\ &&
-{192}\,{\Lx^2}\,{\Ly^2}+{352}\,{\Lx^2}\,{\Ly}-{80}\,{\Lx^2}\,{\pi^2}+{752\over 3}\,{\Lx^2}-{32}\,{\Lx}\,{\Ly}\,{\pi^2}-{176}\,{\Lx}\,{\pi^2}-{384}\,{\zeta_3}\,{\Lx}\nonumber \\ &&
-{96}\,{\Ly^2}\,{\pi^2}-{352}\,{\Ly}\,{\pi^2}+{384}\,{\zeta_3}\,{\Ly}+{56}\,{\pi^4}-{968\over 3}\,{\pi^2}\Biggr )  {}\,{\ttouu}\nonumber \\ &&
+\,{}\Biggl ({}-{192}\,{\Lidx}+{192}\,{\Licx}\,{\Lx}-{96}\,{\Libx}\,{\Lx^2}-{4}\,{\Lx^4}-{32}\,{\Lx^3}\,{\Ly}+{88}\,{\Lx^3}\nonumber \\ & &
+{12}\,{\Lx^2}\,{\Ly^2}-{88}\,{\Lx^2}\,{\Ly}+{48}\,{\Lx^2}\,{\pi^2}-{376\over 3}\,{\Lx^2}-{48}\,{\Lx}\,{\Ly}\,{\pi^2}+{376\over 3}\,{\Lx}\,{\Ly}\nonumber \\ & &
+{64\over 15}\,{\pi^4}+{18}\,{\pi^2}\Biggr ) {\utoss}\nonumber \\ & & {}
+{}\Biggl ({48}\,{\Licx}\,{\Lx}+{144}\,{\Licx}\,{\Ly}+{672}\,{\Licx}-{48}\,{\Libx}\,{\Lx^2}-{672}\,{\Libx}\,{\Lx}+{16}\,{\Lx^4}\nonumber \\ & &
-{32}\,{\Lx^3}\,{\Ly}-{4\over 3}\,{\Lx^3}+{24}\,{\Lx^2}\,{\Ly^2}+{12}\,{\Lx^2}\,{\Ly}-{192}\,{\Lx^2}\,{\pi^2}+{1444\over 3}\,{\Lx^2}+{72}\,{\Lx}\,{\Ly}\,{\pi^2}\nonumber \\ & &
+{80\over 3}\,{\Lx}\,{\Ly}-{624}\,{\Lx}\,{\pi^2}+{80}\,{\Lx}-{288}\,{\zeta_3}\,{\Lx}+{509\over 15}\,{\pi^4}-{707}\,{\pi^2}-{36}-{2800}\,{\zeta_3}\Biggr ) {}\,{\one}\Biggr \} \nonumber \\ & &
+ \Biggl \{ u \leftrightarrow t \Biggr \},
\end{eqnarray}
\begin{eqnarray}
\label{eq:finiteC}
{C }&=&{\Biggl \{ }\Biggl ({}-{24}\,{\Lidx}+{24}\,{\Lidy}+{112}\,{\Lidz}-{44}\,{\Licx}\,{\Lx}+{56}\,{\Licx}\,{\Ly}+{74\over 3}\,{\Licx}\nonumber \\ &&
-{56}\,{\Licy}\,{\Lx}+{44}\,{\Licy}\,{\Ly}-{22}\,{\Licy}-{74\over 3}\,{\Libx}\,{\Lx}+{32\over 3}\,{\Libx}\,{\pi^2}+{22}\,{\Liby}\,{\Ly}\nonumber \\ &&
+{25\over 4}\,{\Lx^4}-{26}\,{\Lx^3}\,{\Ly}+{4}\,{\Lx^3}+{14}\,{\Lx^2}\,{\Ly^2}-{37\over 3}\,{\Lx^2}\,{\Ly}+{7}\,{\Lx^2}\,{\pi^2}+{27\over 2}\,{\Lx^2}+{5}\,{\Ls}\,{\Lx^2}\nonumber \\ &&
+{22\over 3}\,{\Lx}\,{\Ly^3}+{11}\,{\Lx}\,{\Ly^2}-{4}\,{\Lx}\,{\Ly}\,{\pi^2}-{11}\,{\Lx}\,{\Ly}+{31\over 6}\,{\Lx}\,{\pi^2}+{12}\,{\zeta_3}\,{\Lx}-{637\over 27}\,{\Lx}-{26\over 3}\,{\Ls}\,{\Lx}\nonumber \\ &&
-{19\over 12}\,{\Ly^4}-{16\over 9}\,{\Ly^3}+{7\over 3}\,{\Ly^2}\,{\pi^2}-{221\over 18}\,{\Ly^2}-{7\over 3}\,{\Ls}\,{\Ly^2}-{25\over 6}\,{\Ly}\,{\pi^2}+{175\over 9}\,{\Ly}-{12}\,{\zeta_3}\,{\Ly}\nonumber \\ &&
-{98\over 9}\,{\Ls}\,{\Ly}+{1\over 5}\,{\pi^4}+{2\over 9}\,{\Ls}\,{\pi^2}+{203\over 54}\,{\pi^2}-{4\over 9}\,{\zeta_3}-{88\over 9}\,{\Ls^2}+{4849\over 162}-{386\over 27}\,{\Ls}\Biggr ) {}\,{\ttoss}\nonumber \\ &&
+{}\Biggl ({224}\,{\Lidx}+{48}\,{\Lidy}-{48}\,{\Lidz}-{88}\,{\Licx}\,{\Lx}-{24}\,{\Licx}\,{\Ly}+{124\over 3}\,{\Licx}\nonumber \\ &&
+{24}\,{\Licy}\,{\Lx}-{48}\,{\Licy}\,{\Ly}+{280\over 3}\,{\Licy}-{124\over 3}\,{\Libx}\,{\Lx}+{64\over 3}\,{\Libx}\,{\pi^2}\nonumber \\ &&
-{280\over 3}\,{\Liby}\,{\Ly}-{31\over 6}\,{\Lx^4}+{6}\,{\Lx^3}\,{\Ly}-{4\over 3}\,{\Lx^3}-{3}\,{\Lx^2}\,{\Ly^2}-{56\over 3}\,{\Lx^2}\,{\Ly}-{55\over 3}\,{\Lx^2}\,{\pi^2}-{2}\,{\Ls}\,{\Lx^2}\nonumber \\ &&
-{70\over 3}\,{\Lx^2}-{6}\,{\Lx}\,{\Ly^3}-{26}\,{\Lx}\,{\Ly^2}-{2\over 3}\,{\Lx}\,{\Ly}\,{\pi^2}+{4}\,{\Ls}\,{\Lx}\,{\Ly}+{148\over 3}\,{\Lx}\,{\Ly}-{22\over 3}\,{\Lx}\,{\pi^2}\nonumber \\ &&
-{124\over 3}\,{\Ls}\,{\Lx}+{938\over 27}\,{\Lx}+{64}\,{\zeta_3}\,{\Lx}+{32\over 9}\,{\Ly^3}-{3}\,{\Ly^2}\,{\pi^2}+{32\over 3}\,{\Ls}\,{\Ly^2}-{4\over 9}\,{\Ly}\,{\pi^2}-{1096\over 27}\,{\Ly}\nonumber \\ &&
+{24}\,{\zeta_3}\,{\Ly}-{829\over 90}\,{\pi^4}-{10\over 9}\,{\Ls}\,{\pi^2}-{356\over 27}\,{\pi^2}-{352\over 9}\,{\Ls^2}-{1544\over 27}\,{\Ls}-{388\over 9}\,{\zeta_3}+{9698\over 81}\Biggr ) {}\,{\tou}\nonumber \\ & & 
+{}\Biggl ({}-{16\over 3}\,{\Licx}+{16\over 3}\,{\Libx}\,{\Lx}+{40\over 9}\,{\Lx^3}-{16\over 3}\,{\Lx^2}\,{\Ly}+{22\over 9}\,{\Lx^2}+{16\over 3}\,{\Ls}\,{\Lx^2}-{32\over 3}\,{\Ls}\,{\Lx}\,{\Ly}\nonumber \\ & & 
+{14\over 3}\,{\Lx}\,{\pi^2}-{224\over 27}\,{\Lx}-{352\over 9}\,{\Ls}\,{\Lx}+{32\over 9}\,{\Ly^3}+{32\over 3}\,{\Ls}\,{\Ly^2}+{116\over 9}\,{\Ly}\,{\pi^2}-{1096\over 27}\,{\Ly}+{56\over 9}\,{\Ls}\,{\pi^2}\nonumber \\ & & 
+{340\over 27}\,{\pi^2}-{1544\over 27}\,{\Ls}+{9698\over 81}+{32\over 9}\,{\zeta_3}-{352\over 9}\,{\Ls^2}\Biggr )  {}\,{\ttouu}\nonumber \\ & & 
+\,{}\Biggl ({}-{8\over 3}\,{\Licx}+{8\over 3}\,{\Libx}\,{\Lx}-{20\over 9}\,{\Lx^3}+{4\over 3}\,{\Lx^2}\,{\Ly}-{11\over 9}\,{\Lx^2}-{8\over 3}\,{\Ls}\,{\Lx^2}+{11}\,{\Lx}\,{\Ly}-{\Lx}\,{\pi^2}\nonumber \\ & &
+{112\over 27}\,{\Lx}+{176\over 9}\,{\Ls}\,{\Lx}-{2\over 9}\,{\Ls}\,{\pi^2}-{203\over 54}\,{\pi^2}+{88\over 9}\,{\Ls^2}-{4849\over 162}+{386\over 27}\,{\Ls}+{4\over 9}\,{\zeta_3}\Biggr ){\utoss} \nonumber \\ & & {}
+{}\Biggl ({136}\,{\Lidx}-{68}\,{\Licx}\,{\Lx}+{120}\,{\Licx}\,{\Ly}+{206\over 3}\,{\Licx}-{206\over 3}\,{\Libx}\,{\Lx}-{71\over 12}\,{\Lx^4}\nonumber \\ & &
+{14\over 3}\,{\Lx^3}\,{\Ly}-{68\over 9}\,{\Lx^3}+{15}\,{\Lx^2}\,{\Ly^2}+{5\over 3}\,{\Lx^2}\,{\Ly}-{29\over 3}\,{\Lx^2}\,{\pi^2}+{973\over 18}\,{\Lx^2}+{77\over 3}\,{\Ls}\,{\Lx^2}\nonumber \\ & &
-{62\over 3}\,{\Lx}\,{\Ly}\,{\pi^2}-{139\over 6}\,{\Lx}\,{\Ly}-{8}\,{\Ls}\,{\Lx}\,{\Ly}-{317\over 18}\,{\Lx}\,{\pi^2}-{1375\over 27}\,{\Lx}-{626\over 9}\,{\Ls}\,{\Lx}+{4}\,{\zeta_3}\,{\Lx}\nonumber \\ & &
-{47\over 30}\,{\pi^4}+{3799\over 108}\,{\pi^2}+{47\over 9}\,{\Ls}\,{\pi^2}-{484\over 9}\,{\Ls^2}-{2825\over 27}\,{\Ls}+{932\over 9}\,{\zeta_3}+{70025\over 324}\Biggr ) {}\,{\one}\Biggr \} \nonumber \\ & &
+ \Biggl \{ u \leftrightarrow t \Biggr \},
\end{eqnarray}
\begin{eqnarray}
\label{eq:finiteD}
{D }&=&{\Biggl \{ }\Biggl ({24}\,{\Lidx}-{24}\,{\Lidy}+{88}\,{\Lidz}-{52}\,{\Licx}\,{\Lx}+{36}\,{\Licx}\,{\Ly}-{46\over 3}\,{\Licx}\nonumber \\ &&
-{36}\,{\Licy}\,{\Lx}+{52}\,{\Licy}\,{\Ly}+{46\over 3}\,{\Licy}-{4}\,{\Libx}\,{\Lx^2}+{46\over 3}\,{\Libx}\,{\Lx}+{44\over 3}\,{\Libx}\,{\pi^2}\nonumber \\ &&
-{16}\,{\Liby}\,{\Lx}\,{\Ly}+{4}\,{\Liby}\,{\Ly^2}-{46\over 3}\,{\Liby}\,{\Ly}+{79\over 12}\,{\Lx^4}-{82\over 3}\,{\Lx^3}\,{\Ly}+{817\over 18}\,{\Lx^3}+{3}\,{\Lx^2}\,{\Ly^2}\nonumber \\ &&
-{184\over 3}\,{\Lx^2}\,{\Ly}+{13\over 3}\,{\Lx^2}\,{\pi^2}-{545\over 6}\,{\Lx^2}+{38\over 3}\,{\Lx}\,{\Ly^3}+{136\over 3}\,{\Lx}\,{\Ly^2}+{4\over 3}\,{\Lx}\,{\Ly}\,{\pi^2}+{155\over 3}\,{\Lx}\,{\Ly}\nonumber \\ &&
-{10}\,{\Lx}\,{\pi^2}-{32}\,{\zeta_3}\,{\Lx}+{76\over 3}\,{\Lx}-{35\over 12}\,{\Ly^4}-{529\over 18}\,{\Ly^3}+{3}\,{\Ly^2}\,{\pi^2}+{235\over 6}\,{\Ly^2}+{10}\,{\Ly}\,{\pi^2}-{76\over 3}\,{\Ly}\nonumber \\ &&
+{32}\,{\zeta_3}\,{\Ly}-{11\over 30}\,{\pi^4}+{7\over 2}\,{\pi^2}+{8}\,{\zeta_3}+{2}\,{\Ls}-{55\over 6}\Biggr ) {}\,{\ttoss}\nonumber \\ &&
+{}\Biggl ({176}\,{\Lidx}-{48}\,{\Lidy}+{48}\,{\Lidz}-{104}\,{\Licx}\,{\Lx}+{32}\,{\Licx}\,{\Ly}-{92\over 3}\,{\Licx}\nonumber \\ &&
-{32}\,{\Licy}\,{\Lx}+{64}\,{\Licy}\,{\Ly}-{184\over 3}\,{\Licy}-{8}\,{\Libx}\,{\Lx^2}+{92\over 3}\,{\Libx}\,{\Lx}+{160\over 3}\,{\Libx}\,{\pi^2}\nonumber \\ &&
+{16}\,{\Liby}\,{\Lx}\,{\Ly}-{16}\,{\Liby}\,{\Ly^2}+{184\over 3}\,{\Liby}\,{\Ly}-{23\over 6}\,{\Lx^4}-{10}\,{\Lx^3}\,{\Ly}-{385\over 9}\,{\Lx^3}+{19}\,{\Lx^2}\,{\Ly^2}\nonumber \\ &&
+{161\over 3}\,{\Lx^2}\,{\Ly}-{17}\,{\Lx^2}\,{\pi^2}+{80\over 3}\,{\Lx^2}-{14\over 3}\,{\Lx}\,{\Ly^3}-{87}\,{\Lx}\,{\Ly^2}-{26\over 3}\,{\Lx}\,{\Ly}\,{\pi^2}-{260}\,{\Lx}\,{\Ly}\nonumber \\ &&
+{215\over 3}\,{\Lx}\,{\pi^2}-{152\over 3}\,{\Lx}+{168}\,{\zeta_3}\,{\Lx}+{7}\,{\Ly^2}\,{\pi^2}+{545\over 3}\,{\Ly}\,{\pi^2}+{8}\,{\Ly}-{32}\,{\zeta_3}\,{\Ly}-{571\over 90}\,{\pi^4}\nonumber \\ &&
+{742\over 3}\,{\pi^2}+{188\over 3}\,{\zeta_3}-{110\over 3}+{8}\,{\Ls}\Biggr ) {}\,{\tou}\nonumber \\ &&
+{}\Biggl ({32}\,{\Lx^3}-{64}\,{\Lx^2}\,{\Ly}-{310\over 3}\,{\Lx^2}+{32}\,{\Lx}\,{\pi^2}+{64}\,{\Ly}\,{\pi^2}+{8}\,{\Ly}+{352\over 3}\,{\pi^2}+{8}\,{\Ls}\nonumber \\ &&
-{110\over 3}+{32}\,{\zeta_3}\Biggr ) {}\,{\ttouu}\nonumber \\ &&
+\,{}\Biggl ({}-{16}\,{\Lx^3}+{16}\,{\Lx^2}\,{\Ly}+{155\over 3}\,{\Lx^2}-{155\over 3}\,{\Lx}\,{\Ly}-{7\over 2}\,{\pi^2}-{8}\,{\zeta_3}-{2}\,{\Ls}+{55\over 6}\Biggr ) {\utoss}\nonumber \\ & & {}
+{}\Biggl ({64}\,{\Lidx}-{20}\,{\Licx}\,{\Lx}-{108}\,{\Licx}\,{\Ly}-{46}\,{\Licx}-{12}\,{\Libx}\,{\Lx^2}\nonumber \\ & &
+{46}\,{\Libx}\,{\Lx}+{5\over 12}\,{\Lx^4}-{10}\,{\Lx^3}\,{\Ly}-{401\over 18}\,{\Lx^3}-{21\over 2}\,{\Lx^2}\,{\Ly^2}-{34\over 3}\,{\Lx^2}\,{\Ly}-{1\over 3}\,{\Lx^2}\,{\pi^2}\nonumber \\ & &
-{1303\over 6}\,{\Lx^2}-{16\over 3}\,{\Lx}\,{\Ly}\,{\pi^2}-{11\over 6}\,{\Lx}\,{\Ly}+{340\over 3}\,{\Lx}\,{\pi^2}+{104}\,{\zeta_3}\,{\Lx}-{52\over 3}\,{\Lx}-{67\over 20}\,{\pi^4}\nonumber \\ & &
+{2981\over 12}\,{\pi^2}+{11}\,{\Ls}+{1166\over 3}\,{\zeta_3}-{461\over 12}\Biggr ) {}\,{\one}\Biggr \} + \Biggl \{ u \leftrightarrow t \Biggr \},
\end{eqnarray}
\begin{eqnarray}
\label{eq:finiteE}
{E }&=&\Biggl \{
{ }\Biggl ({}-{1\over 3}\,{\Lx^3}-{2\over 3}\,{\Ls}\,{\Lx^2}+{2\over 3}\,{\Lx^2}-{2\over 3}\,{\Lx}\,{\pi^2}+{4\over 3}\,{\Ls}\,{\Lx}-{2\over 3}\,{\Lx}+{1\over 3}\,{\Ly^3}+{2\over 9}\,{\Ly^2}+{2\over 3}\,{\Ls}\,{\Ly^2}\nonumber \\ &&
+{2\over 3}\,{\Ly}\,{\pi^2}+{4\over 9}\,{\Ls}\,{\Ly}+{2\over 3}\,{\Ly}+{2\over 27}\,{\pi^2}+{8\over 9}\,{\Ls^2}\Biggr ) {}\,{\ttoss}\nonumber \\ &&
+{}\Biggl ({2\over 3}\,{\Lx^3}-{2\over 3}\,{\Lx^2}\,{\Ly}+{4\over 3}\,{\Lx^2}+{4\over 3}\,{\Ls}\,{\Lx^2}-{2\over 3}\,{\Lx}\,{\Ly^2}-{8\over 3}\,{\Ls}\,{\Lx}\,{\Ly}+{2\over 3}\,{\Lx}\,{\pi^2}+{8\over 3}\,{\Ls}\,{\Lx}\nonumber \\ &&
+{4\over 3}\,{\Lx}-{2\over 3}\,{\Ly}\,{\pi^2}-{52\over 27}\,{\pi^2}+{4\over 3}\,{\Ls}\,{\pi^2}+{32\over 9}\,{\Ls^2}\Biggr )  {}\,{\tou}\nonumber \\ &&
+{}\Biggl ({16\over 9}\,{\Lx^2}+{32\over 9}\,{\Ls}\,{\Lx}-{40\over 27}\,{\pi^2}+{32\over 9}\,{\Ls^2}\Biggr )  {}\,{\ttouu}\nonumber \\ &&
+\,{}\Biggl ({}-{8\over 9}\,{\Lx^2}-{16\over 9}\,{\Ls}\,{\Lx}-{2\over 27}\,{\pi^2}-{8\over 9}\,{\Ls^2}\Biggr ) {\utoss}\nonumber \\ & & {}
+{}\Biggl ({}-{\Lx^3}-{2}\,{\Ls}\,{\Lx^2}+{26\over 9}\,{\Lx^2}-{2}\,{\Lx}\,{\pi^2}+{10\over 3}\,{\Lx}+{52\over 9}\,{\Ls}\,{\Lx}-{43\over 27}\,{\pi^2}+{44\over 9}\,{\Ls^2}\nonumber \\ & &
+{1\over 2}+{4}\,{\Ls}\Biggr )  {}\,{\one}\Biggr \} + \Biggl \{ u \leftrightarrow
t \Biggr \},
\end{eqnarray}
\begin{eqnarray}
\label{eq:finiteF}
{F }&=&\Biggl \{
{ 2\over 3}\,{}\Biggl ({}-{\Lx}+{\Ly}\Biggr ) \,{}\Biggl ({3}\,{\Lx^2}-{4}\,{\Lx}\,{\Ly}-{14}\,{\Lx}+{3}\,{\Ly^2}-{6}\,{\Ly}+{2}\,{\pi^2}+{4}\Biggr ) {}\,{\ttoss}\nonumber \\ & & 
+{}\Biggl ({4}\,{\Lx^3}-{8\over 3}\,{\Lx^2}\,{\Ly}-{8\over 3}\,{\Lx^2}+{8\over 3}\,{\Lx}\,{\Ly^2}+{80\over 3}\,{\Lx}\,{\Ly}-{4}\,{\Lx}\,{\pi^2}+{16\over 3}\,{\Lx}-{8\over 3}\,{\Ly}\,{\pi^2}-{24}\,{\pi^2}\Biggr ) {}\,{\tou}\nonumber \\ & & 
-{32\over 3}\,{}\Biggl ({}-{\Lx}^2+{\pi}^2\Biggr )  {}\,{\ttouu}
+\,{}\Biggl ({}-{16\over 3}\,{\Lx^2}+{16\over 3}\,{\Lx}\,{\Ly}\Biggr ) {\utoss}\nonumber \\ & & 
{}+{}\Biggl ({2\over 3}\,{\Lx^3}+{2}\,{\Lx^2}\,{\Ly}+{20}\,{\Lx^2}+{4\over
3}\,{\Lx}\,{\Ly}-{16\over 3}\,{\Lx}\,{\pi^2}+{8\over 3}\,{\Lx}-{64\over
3}\,{\pi^2}\Biggr ){}\,{\one}\Biggr \} + \Biggl \{ u \leftrightarrow t \Biggr
\}.\nonumber
\\
\end{eqnarray}

\section{Summary}
\label{sec:conc}

In this paper we presented analytic expressions for the $\O{\as^4}$ QCD
corrections to the $2 \to 2$ gluon-gluon scattering process due to the
interference of the tree-level diagrams with the two-loop graphs in the
\MSbar\ scheme.  Throughout we employed conventional dimensional
regularisation.

The renormalised matrix elements are infrared divergent and contain poles
down to $\O{1/\ep^4}$.  The singularity structure of one- and two-loop
diagrams has been thoroughly studied by Catani~\cite{catani} who provided a
procedure for predicting the infrared behaviour of renormalised amplitudes.
The anticipated pole structure agrees exactly with that obtained by direct
Feynman diagram evaluation.  In fact Catani's method does not determine the
$1/\ep$ poles exactly, but expects that the remaining unpredicted $1/\ep$
poles are non-logarithmic and proportional to constants (colour factors,
$\pi^2$ and $\zeta_3$).  We find that this is indeed the case, and the
constant $H^{(2)}$ is given in Eq.~(\ref{eq:Htwo}). Its origin is in double
emissions from the final state partons.  It is related to that found for
quark-gluon and quark-quark scattering in a straightforward way and therefore
provides a very strong check on the reliability of our results.

The pole structure of the two-loop contribution is described by
Eq.~(\ref{eq:poles}) while analytic formulae for 
the finite part according to the
colour decomposition of Eq.~(\ref{eq:zi}) are given in
Eqs.~(\ref{eq:finiteA}) to~(\ref{eq:finiteF}). 
The one-loop contributions to the
two-loop pole structure are expressed in terms of the one-loop bubble graph in
$D=4-2\ep$ dimensions and the one-loop box graph in $D=6-2\ep$ dimensions 
for which
series expansions around $\epsilon = 0$ are provided in
Appendix~\ref{app:master_int}.

The results presented here, together with those previously computed for
quark-quark scattering~\cite{qqQQ,qqqq,1loopsquare} and quark-gluon
scattering~\cite{qqgg} form a complete set of two-loop hard scattering matrix
elements for parton-parton scattering at $\O{\as^4}$. They are vital
ingredients for the next-to-next-to-leading order predictions for jet cross
sections in hadron-hadron collisions.  However, they are insufficient to make
physical predictions and much work remains to be done.  A major task is to
establish a 
systematic procedure for analytically cancelling the infrared divergences
between the tree-level $2 \to 4$, the one-loop $2 \to 3$ and the $2\to 2$
processes for semi-inclusive jet cross sections.
Recent progress in determining the singular limits of tree-level matrix
elements when two particles are unresolved~\cite{tc,ds} and the soft and
collinear limits of one-loop amplitudes~\cite{sone,coldec2,cone}, together
with the analytic cancellation of the infrared singularities in the somewhat
simpler case of $e^+e^- \to {\rm photon} + {\rm jet}$ at next-to-leading
order~\cite{aude}, suggest that the technical problems will soon be solved
for generic $2 \to 2$ scattering processes.

A further complication is due to initial state radiation. Factorization of
the collinear singularities from the incoming partons requires the evolution
of the parton density functions to be known to an accuracy matching the hard
scattering matrix element.  This entails knowledge of the three-loop
splitting functions.  At three-loop order, the even moments of the splitting
functions are known for the flavour singlet and non-singlet structure
functions $F_2$ and $F_L$ up to $N=12$ while the odd moments up to $N=13$ are
known for $F_3$~\cite{moms1,moms2}. The numerically small $\NF^2$ non-singlet
contribution is also known~\cite{Gra1}. Van Neerven and Vogt have provided
accurate parameterisations of the splitting functions in
$x$-space~\cite{NV,NVplb} which are now starting to be implemented in the
global analyses~\cite{MRS}.

Finally, and most importantly for phenomenological applications, a numerical
implementation of the various contributions must be developed.  The
next-to-leading order programs for three jet production that have already
been written provide a first step in this direction~\cite{trocsanyi,kilgore}.
We are confident that the problem of the numerical cancellation of residual
infrared divergences will soon be addressed thereby enabling the construction
of numerical programs to provide next-to-next-to-leading order QCD estimates
of jet production in hadron collisions.

\section*{Acknowledgements}

M.E.T. acknowledges financial support from CONACyT and the CVCP.  We
gratefully acknowledge the support of the British Council and German Academic
Exchange Service under ARC project 1050.
This work was supported in part by the EU Fourth Framework Programme
`Training and Mobility of Researchers', Network `Quantum Chromodynamics and
the Deep Structure of Elementary Particles', contract FMRX-CT98-0194
(DG-12-MIHT), in part by the U.S.~Department of Energy under
Grant No.~DE-FG02-95ER40896 and in part by the University of Wisconsin
Research Committee with funds granted by the Wisconsin Alumni Research
Foundation.

\appendix
\section{One-loop master integrals}
\label{app:master_int}
In this appendix, we list the expansions for the one-loop box integrals in
$D=6-2\ep$.
We remain in the physical region $s>0$, $u,t < 0$, 
and write coefficients in terms of logarithms and polylogarithms that are
real in this domain.  More precisely, we use the notation of
Eqs.~(\ref{eq:xydef}) and~(\ref{eq:xydef1}) to define the arguments of the
logarithms and 
polylogarithms. The polylogarithms are defined as in
Eq.~(\ref{eq:lidef}).

We find that the box integrals have the expansion
\begin{eqnarray}
\Bfin &=& \frac{ e^{\ep\gamma}
\Gamma  \left(  1+\epsilon \right)  \Gamma  
\left( 1-\epsilon \right) ^2 
 }{ 2s\Gamma  \left( 1-2 \epsilon  \right)   \left( 1-2 \epsilon  \right) } 
 \left(\frac{\mu^2}{s} \right)^{\ep}
  \Biggl\{
 \frac{1}{2}\lq\(\Lx-\Ly\)^2+\pi^2 \rq\nonumber \\
&& 
 +2\ep \lq
 \Licx-\Lx\Libx-\frac{1}{3}\Lx^3-\frac{\pi^2}{2}\Lx \rq
\nonumber \\
&& 
-2\ep^2\Bigg[
\Lidx+\Ly\Licx-\frac{1}{2}\Lx^2\Libx-\frac{1}{8}\Lx^4-\frac{1}{6}\Lx^3\Ly+\frac{1}{4}\Lx^2\Ly^2\nonumber
\\
&&\qquad
\qquad-\frac{\pi^2}{4}\Lx^2-\frac{\pi^2}{3}\Lx\Ly-\frac{\pi^4}{45}\Bigg]
+ ( u \leftrightarrow t) \Biggr\} + \O{\ep^3},
\end{eqnarray}
and
\begin{eqnarray}
\label{eq:boxst}
{\rm Box}^6(s, t)&=&\frac{e^{\ep\gamma}\Gamma(1+\ep) \Gamma(1-\ep)^2}
{2 u\Gamma(1-2\ep)(1-2\ep)}\,\fu  \Biggl\{ \left(\Lx^2 +2 i\pi
\Lx\right)\nonumber \\
&&+\ep \Biggl[
\left(-2\Licx+2 \Lx \Libx  -\frac{2}{3} \Lx^3+2 \Ly \Lx^2-\pi^2 \Lx+2 \zeta_3
\right)\nonumber \\
&& \qquad \qquad +i\pi\left(2 \Libx +4 \Ly \Lx-\Lx^2-\frac{\pi^2}{3}\right)
\Biggr]\nonumber \\
&&+\ep^2 \Bigg[
\Biggl(2\Lidz+2\Lidy-2\Ly \Licx-2\Lx \Licy+(2\Lx\Ly-X^2-\pi^2)\Libx
\nonumber \\
&&\qquad+\frac{1}{3}\Lx^4-\frac{5}{3}\Lx^3\Ly+\frac{3}{2}\Lx^2\Ly^2+\frac{2}{3}\pi^2\Lx^2-2\pi^2\Lx\Ly+2\Ly\zeta_3+\frac{1}{6}\pi^4\Biggr)
\nonumber \\
&&\qquad \qquad + i\pi \biggl(
-2\Licx-2\Licy+2\Ly\Libx+\frac{1}{3}\Lx^3-2\Lx^2\Ly+3\Lx\Ly^2\nonumber \\
&& \qquad \qquad \qquad \qquad -\frac{\pi^2}{3}\Ly+2\zeta_3
\biggr)\Bigg] \Biggr\} + \O{\ep^3}.
\end{eqnarray}
${\rm Box}^6(s,u)$ is obtained from Eq.~(\ref{eq:boxst}) by exchanging $u$ and
$t$.

Finally, the one-loop bubble integral in $D=4-2 \epsilon$ dimensions 
is given by
\begin{equation} 
 \Bubl =\frac{ e^{\ep\gamma}\Gamma  \left(  1+\epsilon \right)  \Gamma  \left( 1-\epsilon \right) ^2 
 }{ \Gamma  \left( 2-2 \epsilon  \right)  \epsilon   } \fs.
\end{equation}


\end{document}